
\input phyzzx


\hsize=6.0in
\vsize=8.9in
\hoffset=0.0in
\voffset=0.0in
\FRONTPAGE
\line{\hfill BROWN-HET-847}
\line{\hfill February 1992}
\vskip1.5truein
\titlestyle{{ALGEBRAIC STRUCTURES AND EIGENSTATES FOR\break
INTEGRABLE COLLECTIVE FIELD THEORIES}\foot{Work supported in
part by the Department of Energy under
contract DE-AC02-76ER03130-Task A}\foot{Work supported by Brown
University Exchange Program P. I. 135}}
\bigskip
\author{Jean AVAN \foot{On leave of absence from LPTHE Paris 6, France}
and Antal JEVICKI}
\centerline{{\it Department of Physics}}
\centerline{{\it Brown University, Providence, RI 02912, USA}}
\bigskip
\abstract
Conditions for the construction of polynomial eigen--operators for the
Hamiltonian of collective string field theories are explored.  Such
eigen--operators arise for only one monomial potential $v(x) = \mu x^2$
in the collective field theory.  They form a $w_{\infty}$--algebra
isomorphic to the algebra of vertex operators in 2d
gravity.    Polynomial potentials of orders only strictly larger or
smaller than 2 have no non--zero--energy polynomial eigen--operators.
This analysis leads us to consider a particular potential $v(x)= \mu
x^2 + g/x^2$.  A Lie algebra of polynomial eigen--operators is then
constructed for this potential. It is a symmetric
2--index Lie algebra, also represented as a sub--algebra of $U (s\ell
(2)).$
\endpage

{\bf\chapter{Introduction}}

Matrix models, \ie $\,\,$ quantum mechanics models with a $N\times N$ matrix
as dynamical variable, were originally introduced as an approach to
non--perturbative aspects of gauge theories (large-N limit of su(N))
[BIPZ,Co].
It was recently realized that they could be viewed as a natural
regularization of string theory (in space--dimension $\leq$2 ) and
thereby
allowed a non--perturbative approach to it.  This approach turned out
to be extremely fruitful and has recently seen a lot of activity
[BK,DS,GM,GMi,GKN].

The collective field method [JS,DJe] was applied to 1--dimensional matrix
models as a natural description of the dynamics of the singlet sector
(eigenvalues of the matrix).  The resulting field theory was
then extensively studied [P,G,K].  Perturbative computations were achieved
[DJR1,2] and found in agreement with results from other approaches
[Mo,DK].  On the other hand it was shown that the collective field theory was
classically [AJ1] and quantum mechanically [AJ2] Liouville--integrable
[Li,Ar],
in the sense that an infinite number of commuting (local) functionals
of the dynamical field $\alpha(x,t)$ could be constructed.  Of crucial
importance in this theory is the existence of a $w_{\infty}$--algebra
[AJ1,2;MPZ,MS,AS,DDMW].  It is given, both
at the classical and quantum level, by  momenta of powers of the field
$\alpha(x,t)$; the $w_{\infty}$ algebraic structure of these
functionals being triggered by the $U(1)\times U(1)$
Kac--Moody algebra structure of the field $\alpha(x,t)$.

An infinite--dimensional Lie--algebra was seen to play the role of a
spectrum--generating algebra for the particular potential $v(x) = \mu
x^2$ [AJ2].  It gave an infinite sequence of discrete states.  In the
collective approach they appear as composite operators.  In the
continuum, conformal theory language, they are the discrete
higher modes of a string in 1+1 dimension [Po,KPo].  The intertwining
$w_{\infty}$-algebra is given in terms of operator products in this
language [Wi,KPo].

Exact eigenstates were also constructed
for the constant background $(v(x) = 0) $ [Je,No].
They are given by character polynomials of $su(N)$, in a way strongly
reminiscent of the free-fermion, infinite Grassmanian approach of the
Kyoto school [JM].  Such a result is not surprising, given the close
relations between matrix models, collective field theory and fermion
theory [GK].

It must be understood that Liouville--integrability as defined above
does not guarantee the existence of action--angle variables, or
more loosely, of variables linearizing the equations of motion (but
not necessarily action--angle stricto sensu - see for instance
[OP,Pe]).  This is only true in
mechanics ( finite number of degrees of freedom) [Li,Ar].  The question is
therefore legitimate, whether physically meaningful angle--type
variables or operators also exist for the collective field theory.
As mentioned, it is already known that an
algebra of polynomial eigen--operators for the collective--field Hamiltonian
exists for a potential $v(x) = \pm x^2$ [AJ2], but this remains yet an
isolated construction.

We want to address here algebraic aspects of the collective field
theory. They are directly related to the question of complete
integrability.  ``Complete Integrability" now means for us the
existence of an infinite algebra of eigen--operators of the original
Hamiltonian and not simply a hierarchy of commuting Hamiltonians.
In the same (formal) way quantum integrable systems are characterized
by the existence of the quantum group structure which in turn leads to
an exact construction of the complete spectrum of quantum states
[SFT,Dr,Ji,F].
We shall see in fact that the algebraic structures arising in our
particular theories are linear, not quadratic, and one can therefore
consider these algebras as the simplest spectrum--generating
structures[Wi,KPo].

Of course we must put restrictions on the type of operators we are
considering, since one can formally construct any kind of
eigen--operator for any given potential, as will appear in the derivation.
First of all, we shall restrict ourselves to collective theories with
polynomial potentials $v$ belonging to ${\bf C}\,(x,x^{-1})$.
We then ask that the eigen--operators be polynomial (or possibly
rational) in the dynamical field $\alpha(x,t)$ and the variable $x$
itself.  The reasons for this choice
are that:

1. the conserved Hamiltonians themselves have this form.

2. $\alpha(x,t)$ is to be identified with the tachyon field,
and we wish in ulterior
constructions to define an action of these operators on the tachyon Fock
space.   The finite powers of $\alpha$ certainly exist as well--
defined operators.  Any additional operator has to be separately
constructed.

3. In particular the physical meaning of an infinite series $\sim \sum c_n
\, \alpha^n$ is anyway
not quite clear:  in fact, strictly speaking, this operator would
not belong to the vector space generated by the $w_{\infty}$--
algebra.  Inclusion of such operators requires an extension of the
original $w_{\infty}$-framework.

4.  It is consistent to ask for polynomial dependance in $x$ since we
are only considering potentials $v(x)$ in $ {\bf C}\,(x,x^{-1})$.

This  restriction will then
naturally reflect on which potentials $v(x)$ do lead to integrable
collective field theories.

The plan of our paper is as follows:\nextline
We first restrict ourselves to monomial potentials in the
matrix models.  We show that only $v(x) = \pm \, x^2$ has polynomial
eigen--operators with non--zero eigenvalues.  The algebra of these
operators is a $w_{\infty}$-algebra.  Its two indices are interpreted
as the energy (= eigenvalue under adjoint action of Hamiltonian) and
momentum $P_{\phi}$ of the eigen--operator.  This operator $P_{\phi}$
describes a translation--like invariance of the equations of motion
[MPZ], but the corresponding quantum number is not conserved under the Lie
bracket.  This algebra is isomorphic to the $(+) $ - algebra of vertex
operators constructed in [Wi,KPo].

We then consider general polynomial potentials.  We show that
potentials containing only monomials of orders larger than 3 or smaller
than 1, do not have polynomial eigen--operators with non--zero
energy.

{}From the previous analysis, we are lead to the construction, on
algebraic grounds, of a polynomial potential with polynomial
eigen--operators:  $v(x) = \mu \, x^2 + {g\over x^2}$.  This
construction has some features of unicity, closely related to
properties of particular elements in the $w_{\infty}$-algebra,
which makes believe that it will be a more
difficult task to construct other integrable polynomial potentials,
although it cannot of course yet be ruled out.  We then establish the
structure of the algebra of polynomial eigen--operators:  it is a 2--indices
symmetric Lie algebra.
It is described in terms of the enveloping algebra of
$s\ell (2)$ as:  $\{ (J_3 )^n \, (J_+ )^{{m\over 2}} \, (J_- )^{b/2} \quad
n \, \epsilon \, N , b \, \epsilon \, N , - b \leq m \leq 3b \}$
where $s\ell (2) = \{ J^+, \, J^-, \, J^3 \}$; such a representation
with three indices is actually redundant.

Before we begin our discussion, let us fix our notations and recall
the essential features of collective string field theory.
It is described by a Hamiltonian:
$$ H = \int dx \, \{ {1\over 2} \, \Pi ,_x  \, \phi (x) \, \Pi ,_x -
{1\over 6} \, \pi^2 \, \phi^3 + (v(x) - \mu ) \phi \}\eqno\eq$$
The string field $\phi (x)$ is the continuum limit of the dynamical
quantity
$$\phi (\{\lambda \} ) = \sum_{\lambda=1}^N \, \delta \bigl(x -
\lambda_i(t)\bigr)\eqno\eq$$
when $\lambda_i(t)$ are the eigenvalues of the matrix field $M(t)$,
dynamical variable of the
corresponding 1--dimensional hermitean matrix model. \nextline
$\Pi (x)$ is its canonical conjugate:
$$\{ \Pi (x) , \, \phi (y) \} = i \, \delta (x-y)\eqno\eq$$
$v(x)$ is the potential function of the matrix model; $\mu$ is the Fermi
momentum, or equivalently a Lagrange multiplier implementing the
normalization condition:
$$\int \, \phi (x) \, dx = N\eqno\eq$$
$\int dx$ is understood as $\oint \, dx$ acting on an analytic
function, as it naturally arises when $H$ is obtained from a canonical
coadjoint construction such as
described in [AJ3].  It is therefore assumed that
$\phi (x)$ is meromorphic in $x$.
$H$, given in (1.1), is in fact the large-N limit of the collective
Hamiltonian.  Lower--order non--local terms also arise, and are
described in [Je].  They will be ignored here, since we set ourselves in
the $N \rightarrow \infty$ limit.

\noindent Introducing the $U(1)\times U(1)$ Kac--Moody current
algebra:
$$\eqalign{ \alpha_{\pm} (x) \equiv \partial_x \Pi\, \pm \pi \Phi
(x) \quad
; \quad & \{ \alpha_{\pm} (x) ,
\alpha_{\pm} (y) \} = \pm 2 i \pi \, \delta ' (x-
y)\cr
& \{ \alpha_+ , \alpha_-\} = 0 }\eqno\eq$$
we define two classical $w_{\infty}$-algebras:
$$h_m^n = \int \, {\alpha_{\pm}^{m-n}\over m-n} \, n^{m-1} \,
dn\eqno\eq$$
with the well-known Poisson bracket relations [Ba1,2]:
$$\bigl\{ h_{m_1}^{n_1} \, ,
\, h_{m_2}^{n_2} \bigr\} = 2i\pi \bigl\{ (m_2-1 ) n_1
- (m_1-1 ) n_2 \bigr\} h_{m_1+m_2-2}^{n_1+n_2}\eqno\eq$$
We assume in principle that $m-n\geq 0$ and $m\geq 1$ , but the
$w_{\infty}$-algebra can be extended to negative powers
using the definition of $\int dx$ as a contour integral $\oint dx$.
We shall in particular allow $m\leq 1$ in order to include potentials
$v(x)$ with negative powers of $x$.  A
central charge $(1+m_1)\,\delta_{m_1 + m_2 +2}$ is then generated in (1.7).

The Hamiltonian $H$ is now written as:
$$H = \int {\alpha_+^3\over 6} - {\alpha_-^3\over 6} + (v(x) - \mu )
(\alpha_+ -\alpha_- ) \,
dx\eqno\eq$$
an element of $w_{\infty} \oplus w_{\infty}$.  The +/ -- decoupling in
both $H$ and the $w_{\infty}$-algebras allows us to consider solely
from now on the + (or --) part in the diagonalizing problem.

The quantum algebra reproduces exactly the classical one.  We introduce
the analogous $H_m^n$ operators:
$$H_m^n = \int {\alpha_{\pm}^{m-n}\over m-n} \, x^{m-1} \, dx$$

\noindent Note that we have no normal--ordering convention.
Introducing it would add central extensions and further linear
deformations to the $w_{\infty}$-algebra, turning it into a
$W_{\infty}$-algebra [PRS].  The non--normal--ordered operators, however, do
not exhibit such deformations of the algebra, because in this case the
reordering terms which generate such deformations exactly cancel by
symmetry (see [AJ2]).

Let us finally recall the form of the commuting Hamiltonians which
guarantee the (weak) Liouville--integrability of (1.8).

\noindent\underbar{{\bf Theorem (1.9)}}:  \quad
{\it The operators}
$h^{(n)} \equiv \int dx \int d\alpha \, (\alpha^2 + v(x))^n ,
n\in N$, {\it commute amongst themselves.  In particular} $H $ {\it in
(1.8) is} $h^{(1)}$.
$$\eqno\eq$$
\par\underbar{{\bf Proof:}}\quad it immediately follows from the
Poisson structure (1.5) -- after part-integration -- for the classical
quantities, and from the exactness of the identification between the
classical Poisson algebra and the quantum Lie algebra, for the quantum
operators\quad {\bf ::}

{\bf\chapter{Monomial Potentials}}

We have restricted ourselves to polynomial eigen--operators (\ie $\,\,$
polynomial functionals of $\alpha(x,t)$).  They belong to the $w_{\infty}$-
algebra defined above, and can be expressed as:
$$\Theta = \sum_{p,q} \, C_p^q \, h_p^q\eqno\eq$$
$p,q$ belonging to a finite set S of integers, and $C_p^q$ being
constant coefficients.

We shall now discuss in an exhaustive way the case of monomial
potentials $v(x) = gx^n, n \in N^* , g \in R$.  The
corresponding Hamiltonian reads:
$$H = \int {\alpha_+^3\over 6} + g x^n \alpha_+ \, + \, {\rm ( \,
minus-term)}\eqno\eq$$
\par The eigenvalue condition $[H, \Theta ] = \epsilon \Theta$ translates
into a recursion relation
$${1\over 2i\pi} \, \epsilon \,
C_p^q = 2p \, C_{p+1}^{q+2} + gn(q-p) \, C_{p-n+1}^{q-
n}\eqno\eq$$
The study of this recursion relation leads us to the

\noindent\underbar{{\bf Theorem 2.4:}}\quad  {\it Polynomial solutions of the
eigenvalue equation} $[H,\Theta ]
= \epsilon \Theta$ {\it with non--vanishing energy
only exist for} $v(x) = \pm x^2$.
$$\eqno\eq$$
\par\underbar{{\bf Proof:}}\quad  The recursion relation (2.3) relates
3 coefficients of the unknown operator $\Theta$.  These coefficients
live on a 2--dimensional integer lattice $(p,q)$ and take non--zero
values only on a finite number of sites.  For $\epsilon \not= 0$, (1.3)
gives the value of $C_p^q$ as a linear combination of two coefficients
sitting on lattice sites separated from $(p,q)$ respectively by
lattice vectors (1,2) and $(n-1 , n)$.  For $n\not= 2$, these two vectors
are non--colinear; one can therefore redefine the reference axis of the lattice
in a way as to have $(p+1, q+2)$ and $(p-n+1, q-n)$ on the same axis,
but not $(p,q)$, namely by setting $\tilde{q} = (n+2 ) p-nq$.

Defining the new lattice--indexation as $P$ (parallel to the
lattice vector $(n+2 , n))$, $Q$(perpendicular to this lattice
vector), (2.3) now expresses values of coefficients $C (P,Q)$ as
linear combination of $C (P' , Q' \not= Q)$.  Since only a finite number
of $C's$ are non--vanishing, there exists a value of $Q$ beyond which
all $C$ vanish.  By recursion, therefore, all $C's$ on the other side
of this limiting line also vanish; hence no solution exists to (2.3)
with  $\epsilon \not=  0$\quad {\bf ::}

For $n=2$, this argument is not valid, since all 3 coefficients are on
the same line.  This argument is not valid either when $\epsilon =0$.
Indeed the commuting hierarchy of Hamiltonians described by Theorem
(1.9) gives us a set of polynomical eigenvalues for any monomial (even
polynomial!) potential.  However, in view of our original purpose in
constructing such algebras of eigen--operators, having only operators
which commute with the Hamiltonian is insufficient to define a quantum
``integrable" system, and we do not intend to address here the question of
finding all operators commuting with $H$,  and their algebraic
structure.

These conclusions are not modified by the inclusion of the central
term $\bigl[ h_p^{p-1} , h_q^{q-1}\bigr]\nextline
 =(p+1)\,\delta_{p+q+2}$ of the
$w_{\infty}$-algebra.  This central term would simply \underbar{add} a
supplementary equation to (2.3) defining the coefficient of {\bf 1}
$\equiv h_0^0$ in $\Theta$, and not \underbar{modify} eq. (2.3).  Theorem
2.4 can therefore be immediately extended to negative--order monomials
$v(x)= x^{-n}$.

We shall now describe the algebra of polynomial
eigen--operators for $v(x) = \pm x^2$.  Some features of this algebra
were already described in [AJ2].

The recursion relation becomes (reabsorbing $2\pi$ into $\epsilon$ ):
$$\epsilon \, C_p^q = 2p \, C_{p+1}^{q+2} + 2 (q-p)\,  C_{p-1}^{q-
2}\eqno\eq$$
We redefine
$$\eqalign { p = K, q & = 2 + 2K-N , {\rm leading\,\, to:}\cr
\epsilon \, C_K^N & =  2K \, C_{K+1}^N + 2 (K+2-N) \, C_{K-1}^N
}\eqno\eq$$
As expected, the recursion relation degenerates into decoupled 1-index
relations for all values of $N$.  Finiteness of the series (2.1)
imposes that (2.6) be consistently truncated at 2 points, namely for
$K\geq K_{max}$ and $K \leq K_{min}$.  It follows from (2.6) that one
has necessarily:
$$K_{min} = 1 \quad\quad K_{max} = N-3\eqno\eq$$
corresponding respectively to monomials in $\alpha \sim \int
{\alpha_+^{N-3}\over N-3} $ and $\int x^{n-4} \alpha_+$.
Normalizing the
coefficient of $\int {\alpha_+^{N-3}\over N-3} $ to be 1,
we obtain from
(2.6) the first terms of the eigen--operator $\Theta$ as:
$$\Theta = H_1^{4-N} + {\epsilon\over 2} \, H_2^{6-N} \, +
\cdots\eqno\eq$$
Once $N$ and $\epsilon$ are given, the operator $\Theta$ can be
recursively constructed from (2.8) and (2.6).

The algebra of such eigen--operators can immediately be
computed:
Define $\Theta_{1,2} \equiv [\Theta_{1,} \Theta_2 ]$
\item{a)} the energy is conserved by application of the Jacobi identity
to $(H, \Theta_{1,} \Theta_2 )$, hence $\Theta_{1,2}$ is an
eigen--operator with energy $\epsilon_1+ \epsilon_2$.
\item{b)} From (2.8) and (2.6) it then follows that:

\noindent\underbar{{\bf Proposition 2.9:}}\quad  $\Theta (N, \epsilon )$
{\it form a Lie algebra defined
to be:}
$$[ \Theta (N_{1,} \epsilon_1) , \Theta (N_2 , \epsilon_2 )] = \bigl\{
{\epsilon_2\over 2} (4 - N_1) - {\epsilon_1\over 2} \, (4-N_2 )\bigr\}
\, \Theta (N_1 + N_2 - 6, \epsilon_1 + \epsilon_2 )$$
$$\eqno\eq$$
This algebraic relation is valid for any value of $N$ and $\epsilon$ ,
independently of the finiteness of the formal series
(2.8) giving $\Theta$.  We now implement the finiteness condition by
first stating two propositions.

\noindent\underbar{{\bf
Proposition 2.10a:}}\quad   $C_K^N$ {\it is
given by a polynomial of order} $K-1$ {\it in} $\epsilon$.

This is obvious from (2.6) and (2.8)\quad {\bf ::}

\noindent\underbar{{\bf
Proposition 2.10b:}}\quad  {\it The truncation relation} $C_K =
0$ {\it for} $K = N-3$ {\it is equivalent to a} $(N-3)$-{\it degree
polynomial equation for}
$\epsilon$.
$$\eqno\eq$$
Indeed, rewriting (2.6) for $K = N-4, N-3$, gives:

\item{\bullet} $\epsilon \, C_{N-4} = 2 (N-4) \, C_{N-3} - 4 C_{N-5}$.
This gives $C_{N-3} (\epsilon )$ as a ($N-4)$ order
polynomial.
\item{\bullet} $\epsilon \, C_{N-3} = -2 C_{N-4}$.  This is a
consistency condition, written as a polynomial of order $N-3$ in
$\epsilon$\quad {\bf ::}

Hence for a fixed $N$, there exists $N-3$ values of $\epsilon$ such
that the recursion relation (2.6) lead to a polynomial eigen--operator
of eigenvalue $\epsilon$.

We now explicitely construct these operators.  Redefining the
canonical dynamical variable $\alpha(x,t)$ as:
$$\tilde{\alpha}_{\pm}\,(x,t) = (\alpha\,(x,t) \pm x )\eqno\eq$$
(the two signs are equally allowed and lead to two different sets of
operators) induces a rewriting of the Hamiltonian in term of
canonically transformed generators $\tilde{H} (\tilde{\alpha})$ of
the $w_{\infty}$-algebra:
$$H = \tilde{H}_1^{-2} \pm 2 \tilde{H}_2^0\eqno\eq$$
depending on the $\pm$ sign in (2.11))

\noindent The operator $H_2^0$ has the unique feature of stabilizing
individually elements of $w_{\infty}$ as $[H_2^0 , H_n^m ] \sim
H_n^m$.  It follows that generators $\tilde{H}_1^{-n}$ are natural
eigen--operators of $H$ with eigenvalues $\pm 2n$ $(n \geq 0$ in order
to have polynomial functionals of $\alpha$).
Since $\tilde{H}_{1(\pm)}^{-n} = \int {(\alpha \pm x)^{n+1}\over n+1 }
$, comparison with (2.8) allows us to identify:
$$\tilde{H}_{1(\pm )}^{-n} = \Theta (N = n+4 , \, \epsilon = \pm 2n)\eqno\eq$$
The algebraic relation (2.9) then leads to:

\noindent\underbar{{\bf Proposition 2.14:}}
$$\bigl[ \tilde{H}_{1(+)}^{(-n)}, \, \tilde{H}_{1 (-)}^{(-m)}
\bigr] \equiv nm \,\, \Theta (n+m+2, 2 (n-m))\eqno\eq$$
This operator is therefore automatically polynomial in $\alpha$.
 Since for each value
of $N = n+m+2$ one has exactly $N-3 = n+m-1$ allowed values of
$\epsilon$, and since $(n-m)$ takes precisely $(n+m-1)$ distinct
values when $n+m$ is fixed and $(n\not= 0 , m\not= 0)$, the set
$\Theta (n+m+2 , 2(n-m))$ completely solves (2.3)

Finally, from (2.14) and (2.9), we end up with:

\noindent\underbar{{\bf Theorem 2.15:}}\quad
{\it The algebra of polynomial eigen--operators for the potential} $v(x) = -
x^2$ {\it is a} $w_{\infty}$-{\it algebra defined by:}
\item{\bullet} $\Theta (n+m+2, \, 2(n-m) ) \equiv B \bigl({(n+m-2\over
2} \, ,{n-m\over 2}\bigr)
\equiv \bigl[ \tilde{H}_{1(+),}^{(-n)} \, \, \tilde{H}_{1(-
)}^{(-m)} \bigr]$
\item{\bullet} $\bigl[ B (J_1 , m_1 ), B (J_2 , m_2 )\bigr] = (J_2 m_1 - J_1
m_2 ) \, B (J_1 + J_2 - 1 , m_1 +
m_2)$
$$\eqno\eq$$
\par $(J,m)$ behave as angular momentum variables  $(J, j_3 )$;
indeed it follows from the
definition in (2.15) that $J \in {1\over 2} N$ and $m = - J, - J+1
\cdots + J$.  The operators $B(J,m)$ are the quantum, continuum version of
the ``angle" variables used in [Pe] to solve exactly the Calogero-
Perelomov integrable mechanics problem [Ca,OP].  A direct proof
of the equivalence of the collective field theory at
$v=x^2$ and the Calogero model is indeed given in [AJ1].

This algebra is remarkably similar to the algebra of ($+$)--type vertex
operators for 2--dimensional gravity introduced in [KPo].
The similitude extends beyond the formal identification
of the Lie brackets, once we describe the physical meaning of the two
indices $J$ and $m$ in (2.15).  $m$ is clearly the energy eigenvalue for
the operator $B(J,m)$.  Interpretation of $J$ requires a first

\noindent\underbar{{\bf Lemma 2.16:}}\quad
{\it The classical equations of motion induced by the Hamiltonian} $H$:
$\alpha ,_t = x - \alpha\alpha , x$ {\it have a translation--like
invariance:} $\delta\alpha = - x \partial_x \alpha + \alpha$.
$$\eqno\eq$$
The proof of this lemma, formulated in [MPZ], is immediate.  This
translation--like invariance acts on the classical $w_{\infty}$-
algebra as follows:

\noindent\underbar{{\bf Proposition 2.17:}}
$$\delta \cdot \int x^{m-1} \, \alpha^{m-n} = (2m-n) \int x^{m-1} \,
\alpha^{m-n}\eqno\eq$$
(obviously follows from partial integration once (2.16) is plugged
into $h_m^n$).  Hence one can formally define the associated
``quantum" operator by its action on $H_m^n$, namely

\noindent\underbar{{\bf Definition 2.18:}}
$$P_\phi \cdot H_m^n = (2m-n)\, h_m^n    \eqno\eq$$
\par It must be emphasized that $P_\phi$ is \underbar{not} generated by the
adjoint action of an operator inside $w_{\infty}$ , since its
eigenvalues are not conserved under the Lie--bracket of $w_{\infty}$.  In
fact, $P_\phi$ loses
a factor 4 under the the Lie bracket.  Although $P_\phi$ is
not an internal morphism of the algebra, it is not surprising that the
eigen--operators of the Hamiltonian (which classically induces the
equations of motion, and therefore generates a flow commuting with the
$P_\phi$ --flow (2.16)) are also eigen--operators of $P_\phi$.

\noindent\underbar{{\bf Proposition 2.19:}}
$$P_{\phi} \cdot \, B(J , m) = 2 (J + 1)\eqno\eq$$
The quantity which is really conserved under the Lie bracket, however, is
$P_{\phi} - 4 = 2(J-1)$.  This is precisely twice the Liouville energy
of the vertex operators in [KPo].  Since the
Liouville mode $\phi$ in 2--dimensional gravity can be understood as a
space--like variable, the Liouville energy is also the eigenvalue of a
space--translation like operator.

We therefore see a one to one equivalence between the collective
theory $w_{\infty}$-
algebra (2.15) and the vertex operator algebra for the Liouville - 2d
gravity formulation.  Let us finally mention that it is proposed to
consider this
$w_{\infty}$-algebra structure to be the fundamental object in a
first--principle approach to string theory [W,AJ3].

{\bf\chapter{Polynomial Potentials.  General Case}}

The problem of classifying all integrable polynomial potentials is much
more involved.  The recursion relation generalizing (2.3) now
contains as many terms as the potential has many monomials, but simple
convexity--type arguments such as the one used to discard all but
$x^2$-monomials, are  available for particular classes of polynomial
potentials.

We now prove the following

\noindent\underbar{{\bf Theorem 3.1:}}\quad   {\it Polynomial potentials of the
form} $v(x) = \sum_{k>2} \, a_k \, x^k$, {\it and} $v(x) = \sum_{k<2} \, b_k
\, x^k$, {\it only have zero--energy polynomial eigen--operators.}
$$\eqno\eq$$
\par\underbar{{\bf Proof:}}\quad   For a general polynomial (allowing
negative powers of $x$) potential of the form $v(x) = \sum_{k\in
S} \, c_k \, x^k,\,k$ belonging to a finite set $S$ in $Z$, the
eigenvalue condition applied to an operator defined by (2.1) reads:
$$\epsilon \, C_p^q = 2p \, C_{p+4}^{q+2} + \sum \, c_k \cdot k (q-p)
\, C_{p-k+1}^{q-k}\eqno\eq$$
generalizing straightforwardly (2.3).  When $\epsilon \not= 0$, this
gives the value of $C_p^q$ at a site $(p,q)$ of the lattice $Z\times
Z$ as a well--defined linear combination of coefficients located at
sites translated by the lattice vectors (1,2) and $(-k+1 , -k)$ for
all $k\in S$.

If $v(x)$ contains only monomials of order $k>2$ or monomials of order
$k<2$, these lattice vectors are included in a sector of angle smaller
than $\pi$, limited by the most external vectors $(1,2)$ and $(-2, -
3)$  (for $k>2)$ and $(1,2)$ and $(0,-1)$ ( for $k<2$).  We now shift the
origin of the lattice to $(p,q)$.

\underbar{For $k>2$} we define two
half--planes with boundary $5p-3q =-1$.  The new origin of the lattice
lies on the strictly positive side $(5p-3q+1>0)$, the extremity of all
translation vectors (i.e. the lattice points defining the value of $C$
at the origin point) lie on the negative side $(5 p - 3q + 1 = -
2k+5\leq 0)$.  Defining now a relabeling of the lattice points, as
$\tilde{Q} = 5p - 3q$, we have
shown that (3.2) gives a linear, well--defined for $\epsilon \not= 0$,
relation between the coefficient $C$ at a given $(p,\tilde{Q_0}$) and
coefficients $C$ with $\tilde{Q} < \tilde{Q_0}$.

A polynomial eigen--operator having only a finite number of non--vanishing
coefficients in (2.1), there exists a $\tilde{Q}_0$ such that all $C$
vanish for $Q<\tilde{Q}_0$, and one $C$ at least is non--zero for
$Q=\tilde{Q}_0$.  However, we see that $C(\tilde{Q}_0 )= \sum q_n C
(Q<\tilde{Q}_0 ) = 0$.  Hence no polynomial eigen--operator exists
with $\epsilon \not= 0$.

\underbar{For $k<2$}, the demonstration runs on similar lines:  the half--plane
boundary is here defined by the equation $3p-q-1=0$\quad {\bf ::}

The theorem does not preclude zero--energy eigen--operators.  Indeed
Theorem (1.9) guarantees the existence of a class of such operators
for any potential $v(x)$, namely the hierarchy Hamiltonians $\int
d\alpha\int dx \, (\alpha^2 + v(x))^n$.

If $v(x)$ contains a term $x^2$, the two translation vectors ($1,2$)
and $(-k+1, -k), k=2$, become colinear.  The translation sector is flat,
and the convexity argument allowing the introduction of the new index
$\tilde{Q}$ is not valid:  the origin of the shifted lattice now {\it
also} lies on the boundary of the half--plane defined by the
externmost vectors $(1,2)$ and $(-1-2)$.

Finally if $v(x)$ contains both terms of order $k>2$ and $k<2$, the
translation--vector sector has an angle $\theta > \pi$ and the convexity
argument collapses as well, the origin of the shifted lattice being
now \underbar{inside} the negative half--plane.

The general study of these potentials, leading to non--convex recursion
relations, is certainly much more involved since we lack the previous
simple geometric arguments.  We are however able to construct at
least one explicit
realization of a finite non--convex recursion relation, using
restricting hypothesis which will considerably simplify the study.

{\bf\chapter{An Integrable potential:  $v(x) = \mu x^2 +
{g\over x^2}$.}}

In order to obtain an integrable potential, we have to break from the
structures of $v(x)$ described in Theorem 3.1.  We now describe what
appears to be the most natural construction of $v(x)$ on such lines.

{\bf 1.  \underbar{Construction of the Potential}}

The mildest assumption in order to get an
integrable potential is to allow an $x^2$--term.  We accordingly formulate:

\noindent\underbar{{\bf Hypothesis 1}} --  The potential $v(x)$
contains a $\mu x^2$
term.

This makes it possible, using the redefinition $\alpha
\rightarrow (\alpha \pm i \sqrt{\mu} x )$, to rewrite
$$\tilde{H} = \tilde{H}_1^{-2} \pm i\sqrt{\mu} \, H_2^0 +
(\bar{v}-{\rm term} )\eqno\eq$$
where $(\bar{v}$--term) means all remaining linear terms in $\alpha$
induced by $\int (v - \mu x^2 )\alpha \, dx$.  In this way, since
$H_2^0$ is the stabilizing element of the $w_{\infty}$-algebra, the
general recursion relation for the coefficients $C_p^q$ in (2.1) loses
one term.

\noindent \underbar{{\bf Hypothesis 2}} -- There exists polynomial
eigen--operators diagonalizing simultaneously
$\bigl(\tilde{H}_1^{-2} + (\bar{v}$--term)$\bigr)$ and $H_2^0$.

Although $H_2^0$ is a stabilizing term, this is a rather strong
restriction, but it will ultimately simplify considerably the problem.
Indeed, in many similar algebraic problems, the best way of
diagonalizing a Hamiltonian is in fact to diagonalize a family of
Hamiltonians, and the quantum inverse scattering method [SFT,F] relies
precisely on such a formulation.  In this particular case, we know
that:

\noindent\underbar{{\bf Proposition 4.2:}}
\quad  $H_2^0$ {\it is diagonalized by
linear combinations of operators in} $w_{\infty}$ {\it with the same}
$n$-{\it index} - (obvious from (1.6)).
$$\eqno\eq$$
\noindent\underbar{{\bf Proposition 4.3:}}
\quad  {\it The operator} $\bigl(\tilde{H}_1^{-2} + \bar{v}$ -
term$\bigr)$ {\it has polynomial eigen--operators} $h_{\pm}^{(n)}$
$$h_{\pm}^{(n)} = \int dx \int d\tilde{\alpha} \,
(\tilde{\alpha}_{\pm}^2 + \bar{v} (x))^n\eqno\eq$$
This is immediately obtained from the fact that this operator
is exactly the collective field theory Hamiltonian for a
potential $\bar{v} = (v - \mu x^2)$, and therefore there exists an
infinite set of commuting Hamiltonians according to Theorem
(1.9).

\noindent\underbar{{\bf Hypothesis 3}} --  These Hamiltonians also diagonalize
$H_2^0$.

This hypothesis, although again restrictive, is fortunately
not empty and will provide us with a (unique) example of
integrable potential.

\noindent\underbar{{\bf Proposition 4.4:}}  \quad
{\it The potential} $\bar{v}(x) = g/x^2$ {\it is the only one which
fulfills Hypothesis 3.}
$$\eqno\eq$$
\par\underbar{{\bf Proof:}}\quad  In order for the Hamiltonians $h^{(n)}$
to diagonalize $H_2^0$, they must be expanded as elements of the
$w_{\infty}$-algebra (1.6) with the same $n$-index.
Since $n$ is equal, up to a constant, to
(degree of $x$) minus (degree of $\alpha$), it follows that
$\bar{v}(x)$ must be a monomial of order -2 to fulfill
Hypothesis 3\quad {\bf ::}

The Hamiltonians $h_{\pm}^{(n)}$ have an eigenvalue $ \pm
\sqrt{\mu}2n \pi$ under the complete Hamiltonian for $v = \mu x^2 +
g/x^2$.  We
shall usually disregard the factor $2\pi\sqrt{\mu}$ in the energy, except
when explicitely required for needs of a demonstration.

The use of the operators $h^{(n)}$ in the  last hypothesis is actually
almost unavoidable.  If we do not want to solve directly the
recursion relation (3.2) for the potential $\bar{v}(x)$, we have to
use the eigen--operators known originally from Theorem (1.9).
Moreover, in the
(in principle) simplest case when $\bar{v} (x)$ is a monomial $\not=
x^2$, Theorem (2.4) precludes the existence of \underbar{any} other
diagonalizing polynomial operator.  So does Theorem (3.1) when
$\bar{v}(x)$ is a polynomial of order 3 and more, or 1 and less.

We see therefore that any deviation from this set of three
hypothesis leads us necessarily to directly tackling the general,
non--convex recursion relation (3.2) without any further help from the
algebraic structure $w_{\infty}$.  In this sense, the potential $v(x)
= \mu x^2 + {g\over x^2}$ has unique features.

We now describe a naturally generated algebra of polynomial
eigen--operators for $v(x) = \mu x^2 + g/x^2$.
Although it is not proved to be the full algebra of such
operators, it exhibits nevertheless interesting structures which makes it
worthy of investigation.

{\bf 2. \underbar{Algebra associated to $v = \mu x^2 + g/x^2$}}

The algebra of eigen--operators which we are going to construct is
formally obtained by taking successive commutators of the Hamiltonians
$h_{\pm}^{(n)}$ with themselves and with the successively generated
operators.  Our problem is to describe explicitely what this
enveloping algebra is .  We shall consider for the moment that
the $w_{\infty}$-algebra of generators $\oint x^{m-1} \alpha^{m-n}$
has no central terms.  We have previously mentioned that such terms
actually appear in commutators $[h_m^{m-1} , h_n^{n-1}]$ when $n+m=-
2$,  and we shall study their
effect later.

We now define candidate eigen--operators as polynomials in $\alpha$ of
the form:
$$B = \sum_{n=1}^{N\cdot} \, \int dx \, f_n (x) \, {\alpha^n\over
n}\eqno\eq$$

The eigenvalue condition $[H,B] = \epsilon B$ is equivalent to the
recursion relation:
$$2n (-\mu x^2 + {g\over x^2} ) f_{n+1} =  \epsilon x f_n - 2x
\partial_x f_{n-1}\eqno\eq$$

This relation is simply the identification of the functional
coefficients of $\alpha^n$ in the eigenvalue condition.  The following
proposition is now obvious:

\noindent\underbar{{\bf Proposition 4.7:}}\quad  {\it The giving of}
$f_1$ {\it and} $\epsilon$ {\it is necessary and sufficient to define
the operator} $B$.
$$\eqno\eq$$
\par This now helps us to write the first orders in $\alpha^n$ of $B$:
$$B (\epsilon ) = \int dx \, f_1 (x) \, \alpha(x) \quad + \epsilon
\int dx \cdot x \, {f_1(x)\over -\mu x^2 + g/x^2} \, {\alpha^2
(x)\over 2} \, + \cdots\eqno\eq$$
We now assume a particular form for $f_1 (x)$.  We already have an
example given by the Hamiltonians $h_{\pm}^{(n)}$ in Proposition (4.3), for
which:
$$f_1 [h_{\pm}^{(n)}] = (-\mu x^2 + g/x^2 )^n\eqno\eq$$
In fact, the most general form of
$f_1$ for eigen--operators obtained from $h_{\pm}^{(n)}$ is:
$$f_1 =  ( -\mu x^2 + g/x^2 )^b \, (\mu x^2 + g/x^2
)^a ; \,\, a,b,\in N\eqno\eq$$
This follows from

\noindent\underbar{{\bf Proposition 4.11:}}\quad  {\it The eigen--
operators} $B^{(a,b)} (\epsilon )$ {\it defined under Prop. 4.7 by the
energy} $\epsilon$ {\it and an initial function} $f_1$ {\it of the form
(4.10), form a closed Lie algebra.
This  algebra reads:}
$$\eqalign{\bigl[ B^{(a_1,b_1)}(\epsilon_1) , B^{(a_2 ,
b_1)}(\epsilon_2) \bigr]
& = - (a_2 \epsilon_1 - a_1 \epsilon_2 ) B^{(a_1 + a_2 - 1 , b_1 + b_2
)}(\epsilon_1 +\epsilon_2)\cr
& \, - (b_2 \epsilon_1 - b_1 \epsilon_2 ) B^{(a_1 + a_2 + 1 , b_1 + b_1
- 2)}(\epsilon_1 + \epsilon_2)}\eqno\eq$$

\underbar{{\bf Proof:}}

First of all, it follows from applying the Jacobi identity to the
operators $H, B_1$ and $B_2$, that the $l.h.s.$ is an eigen--operator
with eigenvalue $\epsilon_1 + \epsilon_2$.

Now in order to apply Proposition (4.7), we must compute the linear
term in $\alpha$ on the $l.h.s.$ of (4.10).  For two general operators
of the form (4.5) respectively defined as:
$$B_1 = \int f_1 \, \alpha \, + f_2 \, {\alpha^2\over 2} + \cdots
\quad\quad B_2 = \int g_1\, \alpha \, + g_2 \, {\alpha^2\over 2} +
\cdots\eqno\eq$$
the commutation relation (1.5) for $\alpha$ implies that the linear term of
$[B_1, B_2 ]$ reads:
$$ \bigl( - \partial_x \, f_1 \cdot g_2 ' + \partial_x g_1\cdot f_2 '
\bigr)\eqno\eq$$
Inserting now (4.10) and (4.8) into (4.13) and using the fact that $x
\partial_x (\mu x^2 \pm \, g/x^2) = \mu x^2 \mp \, g/x^2$ leads us
immediately to (4.11)\quad {\bf ::}

We now need to establish the set of values for $a, b$ and $\epsilon$
obtained by the repeated action of $h_{\pm}^{(n)} \equiv B^{(0,n)}
(\pm n)$.  In the first place, commutators $[h_+, h_-]$ and $[h_- ,h_-]$
obviously vanish since $h_+$ and $h_-$ are canonical transformations
of original commuting Hamiltonians for $g/x^2 = v(x)$.  We then prove

\noindent\underbar{{\bf Lemma 4.14:}}
$$\eqalign{\bigl\{ [ B^{(0,n_1 )} (+n_1 ) & ,
B^{0,n_2 )} (-n_2 ) ] , n_1 , n_2 \in N \bigr\}\cr
& \equiv \bigl\{ B^{(1,b)}
(\epsilon ), b \in N, \epsilon \in N, \vert \epsilon \vert \leq b , \,
\epsilon = b,b-2\cdots - b\bigr\} }\eqno\eq$$

\underbar{{\bf Proof:}}\quad  From (4.11), these commutators only
generate $B^{(1,n_1 + n_2 - 2)} \, (n_1 - n_2 )$ for $n_1 , n_2$
strictly positive integers.  For a fixed $b = n_1 + n_2 - 2 \geq 0$,
$n_1$ can take all values from $1$ to $b+1$ and respectively $n_2$
goes from $b+1$ to $1$.  Hence $\epsilon = n_1 - n_2$ goes from $b
$ to $b-2 \cdots$ to $-b$ \quad {\bf ::}

This leads us
to consider the set $Q = \{ (b, \epsilon) \in Z \times N, b \geq \vert \epsilon
\vert , \epsilon = b , b-2 \cdots -b \}$.  It is a square sublattice
of $Z
\times Z$, limited by $\epsilon = \pm b$ with lattice spacing 2.  We
now prove

\noindent\underbar{{\bf Lemma 4.15:}}  \quad
{\it For any} $a_1, a_2 \in N$, {\it the
commutator} $[B^{(a_1 , b_1)} (\epsilon_1 ) , B^{(a_2 ,
b_2)} (\epsilon_2)]$, {\it where} $(b_1 , \epsilon_1 )$ {\it and}
$(b_2 , \epsilon_2 )$ {\it belong to}  $Q,$ {\it yields
only} $B$-{\it operators with} $(b, \epsilon)$  {\it inside} $Q$.
$$\eqno\eq$$
\par\underbar{{\bf Proof:}}\quad   Consider the commutation relation
(4.11).  Such a commutator yields two terms:

1.) one has $a = a_1 + a_2 - 1; (b,\epsilon) = (b_1 ,
\epsilon_1 ) + (b_2 ,
\epsilon_2)$.  The lattice $Q$ is closed under addition in $Z\times Z$
since: \nextline
$\vert \epsilon_1 + \epsilon_2 \vert \le \vert \epsilon_1 \vert +
\vert \epsilon_2 \vert \le b_1 + b_2 \quad \forall b_1 , b_2,
\epsilon_1 , \epsilon_2$ in $Q$\nextline
Hence the first term in the commutator is inside $Q$.

2.)  The other term has $a = a_1 + a_2 +1 $; $(b,\epsilon ) =
(b_1 , \epsilon_1 ) + (b_2 - 2 , \epsilon_2 )$.  The closure argument
applies here unless $\epsilon_2 = \pm b_2$, and since one also has
$(b, \epsilon ) = (b_1 - 2 , \epsilon _1 ) + (b_2 , \epsilon_2),$ it
again applies unless $\epsilon_1 = \pm b_1$.  Two different cases are
yet to be discussed:

a)  $\epsilon_1 = \pm b_1 , \epsilon_2 = \pm b_2$:  in this case the
coefficient in front of the term vanishes (see (4.11)).
\item{b)} $\epsilon_1 = \pm b_1 , \epsilon_2 = \pm b_2$.  One can then
rewrites, respectively:\nextline
either $(b,\epsilon ) = (b_1 , b_1 - 2) + (b_2 - 2 , - b_2 + 2) \quad
\in Q$ by closure property.\nextline
or $(b,\epsilon ) = (b_1 , - b_1 + 2) + (b_2 - 2, b_2 - 2) \quad
\in \, Q$ by closure property.

This demonstration holds unless
$b_1 = 0 = \epsilon_1 $;  then the coefficient in (4.11) vanishes
anyway\quad {\bf ::}

\noindent The natural counterpart of (4.15) is:

\noindent\underbar{{\bf Lemma 4.16:}}\quad  {\it The repeated adjoint
action of} $B^{(0,b)}
(\epsilon), (b,\epsilon )\in Q$, {\it generates all} $\bigl\{ B^{(a,b)}
(\epsilon ),\, a \in N, (b, \epsilon ) \in Q \bigr\}$.
$$\eqno\eq$$
\par\underbar{{\bf Proof:}}\quad
Assume we have constructed $B^{(0,b)} (\epsilon
), (b,\epsilon) \in Q$ (we shall soon prove this assumption).  Lemma
(4.16) now follows by recursion:
\item{1)} $a = 0$ is true by assumption.
\item{2)} Once (4.16) is valid for $a$ $\in N$ up to $a_0$, we act by
$B^{(0,b_0 )} (\epsilon_0 )$ on  $B^{(1,b_1)} (\epsilon_1)$.
{}From (4.11) we get two terms
\item{a)}  $a_0 \epsilon \cdot B^{(a_0-1, b_0+b_1)} (\epsilon_1 +
\epsilon_0):$  already constructed by recursion hypothesis.
\item{b)} $(b_1\epsilon_0 - b_0 \epsilon_1 ) \, B^{(a_0 + 1 ,  b_0 +  b_1 - 2)}
(\epsilon_1 + \epsilon_0 ) \simeq B^{(a_0 +1,b)} (\epsilon)$

To get any operator with non--zero energy, use $\epsilon_0 = 0 , b_0 =
b+2 - \vert \epsilon \vert \not= 0$ (automatically since $(b,\epsilon
) \in Q), b_1 = \vert \epsilon_1 \vert = \vert \epsilon \vert \not= 0$
and thus $(b_1 \epsilon_0 - b_0 \epsilon ) \not= 0$.  To get
$\epsilon = 0$, use $b_0 = b_1 = {b\over 2} + 1$ (for $\epsilon = 0,
b$ is necessarily even--positive, in $Q$) and $\epsilon_0 = -
\epsilon_1 = 1 $\quad {\bf ::}

We have left as an assumption the existence of
$B^{(0,b)} (\epsilon )$.  We now prove:

\noindent\underbar{{\bf Lemma 4.17:}}\quad   $B^{(0,b)} (\epsilon ) ,
(b,\epsilon ) \in Q, \epsilon \not= 0$, {\it are polynomial
eigen--operators generated by} $\bigl\{ B^{(0,n)} (\pm n) \bigr\}$.
$$\eqno\eq$$
\par\underbar{{\bf Proof:}}\quad  Lemma (4.14) proved directly the existence of
$B^{(a=1)}$ for $(b,\epsilon )$ in $Q$.  We now
apply again $B^{(0,1)} (\pm 1)$; according to (4.11):
$$\eqalign{\bigl[ B^{(0,1)} (\pm 1) , B^{(1,b)} (\epsilon )\bigr]  =  & \pm 1
B^{(0,b+1)} (\epsilon\pm 1)\cr
& + \bigl(\epsilon -b (\pm 1)\bigr) \, B^{(2,b+1-2)} (\epsilon \pm
1)}\eqno\eq$$
Hence we explicitely construct:
$$\eqalign{ & - B^{(0,b+1)} (\epsilon + 1) + (\epsilon - b) \, B^{(2,b-1)}
(\epsilon + 1) \equiv B^+ (\epsilon )\cr
& - B^{(0,B+1)} (\epsilon - 1) + (-\epsilon -b) \, B^{(2, b-1)}
(\epsilon - 1) \equiv B^-(\epsilon )}$$
For $\epsilon \not= b$, let us consider $B^+ (\epsilon ) - B^-
(\epsilon + 2)$.  We get,
$$B^+ (\epsilon ) - B^- B^- (\epsilon + 2)  = (\epsilon + 1)
\, B^{(2,b-1)} (\epsilon + 1)\eqno\eq$$
$$B^{(0,b+1)} (\epsilon + 1)  = - B^+ (\epsilon ) + {\epsilon - b\over
\epsilon + 1} \, \bigl( B^+ (\epsilon ) - B^- (\epsilon +
2)\bigr)\eqno\eq$$
Hence for all values of $\epsilon +1$ except $\pm (b+1)$ and $0$,
$B^{(0,b+1)} (\epsilon + 1)$ is obtained as a linear combination of
successive commutators of $B^{(0,n)} (\pm n)$.  Finally we have
identified in (4.8) the Hamiltonians $h_{\pm}^{(n)}$
as the eigen-operators $B^{(0,n)} (\pm n)$\quad {\bf ::}
\noindent
We now prove directly the

\noindent\underbar{{\bf Lemma 4.21:}}\quad  {\it Operators} $B^{(0,n)}
(0)$ {\it are polynomial eigen--operators of} $H$.
$$\eqno\eq$$
\par\underbar{{\bf Proof:}}\quad The recursion relation (4.6) for zero-
eigenvalued operators reads:
$$2n (-\mu x^2 + {g\over x^2}) \, f_{n+1} = - 2x \,\partial_x \, f_{n-
1} \eqno\eq$$
Application of the operator $\bigl[ {x\over -\mu x^2 + g/x^2} \,
\partial_x \bigr]$ on a function of the form $\bigl( - \mu x^2 + g/x^2
+\bigr)^b$\nextline $\bigl( \mu x^2 + g/x^2 \bigr)^a$ leads to
$b \bigl( - \mu x^2 + {g\over x^2} \bigr)^{b-2} \, \bigl( \mu x^2
+ {g\over x^2} \bigr)^a + a \quad\quad  \bigl( -\mu x^2 +
{g\over x^2}\bigr)^b \, \bigl( \mu x^2
+ {g\over x^2} \bigr)^{a-1}$.  Hence it decreases the global degree by
$1$ without creating negative degrees if $b$ initially is even.
Successive applications of (4.22) on an initial function $f_1$
with $b=0$ leads to $0$ after $(a+1)$ steps.  Therefore the recursion
relation (4.22) leads to a polynomial operator $B^{(0,n)} (0)$\quad {\bf ::}

We can now state the major

\noindent\underbar{{\bf Theorem 4.23:}} \quad  {\it The operators}
$\bigl\{ B^{(a,b)}
(\epsilon ); a \in N; (b,\epsilon ) \in Q\bigr\}$ {\it form a closed algebra of
polynomial eigen--operators for the potential} $\mu x^2 + g/x^2$.
{\it The algebra structure is:}
$$\eqalign{\bigl[ B^{(a_1 , b_1)} (\epsilon_1 ), B^{(a_2 , b_2 )} (\epsilon_2
)\bigr] & = (a_1 \epsilon_2 - a_2 \epsilon_1 ) B^{(a_1 +a_2 -1 , b_1 +
b_2 )} (\epsilon_1 + \epsilon_2 )\cr
& + (b_1 \epsilon_2 - b_2 \epsilon_1 ) B^{(a_1 + a_2 +1, b_1 + b_2 -
2)} (\epsilon_1 + \epsilon_2 ) }\eqno\eq$$

\underbar{{\bf Proof:}}  This follows obviously from (4.17) and
(4.21), when $a_1 = 0 $; and from (4.16) recursively, when $a_1
\not= 0$\quad {\bf ::}

The commuting Hamiltonians constructed in Theorem (1.9) are here
$B^{(n,0)} (0)$.

\noindent Now the 3-index representation is in fact very much
redundant, although it was maintained until now for
practical purposes.  From (4.9) and (4.7) one has in fact

\noindent\underbar{{\bf Proposition 4.24:}} \quad  {\it For all
allowed values of} $a,b,\epsilon$,
$$B^{(a+2 , b)} (\epsilon ) - B^{(a,b+2)} (\epsilon ) = 4\mu g \,
B^{(a,b)} (\epsilon)\eqno\eq$$
It follows that the values of $a$ are actually reduced to $a = 0, 1$.
The algebra (4.23) reduces to a 2-index symmetric Lie algebra:

\noindent\underbar{{\bf Corollary 4.25:}}\quad  {\it The set of linearly
independant eigen--operators can be chosen as} $\bigl\{ B^{(0,b)}
(\epsilon ), B^{(1,b)} (\epsilon ) , (b, \epsilon ) \in Q \bigr\}$.
{\it The algebra reads:}
$$\eqalign{ \bigl[ B^0 (b_1 , \epsilon_1 ), B^0 (b_2 , \epsilon_2
)\bigr] & = (b_1 \epsilon_2 - b_2 \epsilon_1 ) B^1 (b_1 + b_2 - 2 ,
\epsilon_1 + \epsilon_2 )\cr
\bigl[ B^0 (b_1 , \epsilon_1 ), B^1 (b_2 , \epsilon_2 ) \bigr] & =
(b_1 \epsilon_2 - \epsilon_1 (b_2 + 1) ) B^0 (b_1 + b_2 , \epsilon_1 +
\epsilon_2 ) \cr
& + 4\mu g (b_1 \epsilon_2 - \epsilon_1 b_2 ) B^0 (b_1 + b_2) - 2 ,
\epsilon_1 + \epsilon_1 )\cr
\bigl[ B^1 (b_1 , \epsilon_1 ), B^1 (b_2 , \epsilon_2 )\bigr] & =
\bigl( (b_1 + 1) \epsilon_2 - \epsilon_1 (b_2 + 1)\bigr) B^1 (b_1 +
b_2 , \epsilon_2 + \epsilon_2 )\cr
& + 4\mu g (b_1 \epsilon_2 - b_1
\epsilon_1 ) B^1 (b_1 + b_2 - 2 , \epsilon_1 + \epsilon_2 )
b_2} \eqno\eq$$
The above algebra generalizes the $w_{\infty}$-algebra of the
oscillator potential.  It will be interesting to address the
question of a physical interpretation of this algebra in terms
of discrete states of some 2--dimensional theory.

{\bf 3.  \underbar{Effect of Central Terms}}

We have now constructed an algebra of eigen--operators defined by the
commutation relations (4.25).  The addition of the correct central terms
in the $w_{\infty}$-algebra does not modify sensibly this
demonstration.  Such terms arise only when computing
commutators of linear terms in $\alpha$.  In particular, operators
defined as in (4.5) get a central term $\sim \oint f_1 \, \partial_x
\, g_1 $ in their Lie bracket.  The induced changes are as
follow:
\item{(1)}  in (4.5), $B$ acquires a term $f_0 \cdot {\bf 1}$
\item{(2)}  in (4.6), one must add a recursion relation for $f_0$:
$$\epsilon \, f_0 = \oint \, f_1 \cdot (\mu x - g/x^3 )\, .\eqno\eq$$
For $\epsilon = 0$, however, we prove that:

\noindent\underbar{{\bf Proposition 4.27:}}
$$\oint f_1 (\mu x - g/x^3 ) dx = 0\eqno\eq$$
\par\underbar{{\bf Proof:}}\quad   Inside our
initial algebra (4.11), $\epsilon =
0$ implies $b$ even.  Hence
$$\eqalign{\oint \, f_1 (\mu x - g/x^3 ) dx & = \oint (\mu x^2 - g/x^2
)^b \, (\mu x^2 + g/x^2 )^a (\mu x - g/x^3 ) dx\cr
{\rm (part-integration)} & = \oint {b\over q} (\mu x^2 - g/x^2 )^{b-1}
(\mu x^2 + g/x^2 )^{a+1} (\mu x + g/x^3)\cr
& = \oint {b\over a} (\mu x^2 - g/x^2 )^{b-2} (\mu x^2 + g/x^2 )^{2+2}
(\mu x - g/x^3 ) dx \cr
& = \cdots 0\,\, {\rm after\,\,reaching}\,\, b=0\quad {\bf ::}}$$
Hence (4.26) determines consistently $f_0$ for $\epsilon \not= 0$.
Proposition (4.7) is not modified.  Proposition (4.11) could only be
modified by a central extension, but the Jacobi identity applied to
$B_1 , B_2$ and $H$ implies:
$$\bigl[ H, [B_1 , B_2 ]\bigr] = (E_1 + E_2) \, [B_1 , B_2 ]\eqno\eq$$
since $B_1 , B_2$ are now exact eigenstates of $H$; from (4.25) it
appears that no extra central term can be generated in the {\it exact}
commutators of the {\it exact} eigen--operators.  Since all further
derivations follow from (4.11), we conclude that the central extension of the
$w_{\infty}$-algebra does not modify the 2-index symmetric algebra of
eigen--operators for $\mu x^2 + g/x^2$.

{\bf 4. {\underbar{Relation with $S\ell (2)$}}}

It is easier to work here with the 3-index redundant representation.
This 3-index algebra (4.23) can be interpreted as a subalgebra of the
enveloping algebra of $s\ell (2)$, at least at the classical level, and
allowing negative integer powers of the generators.  Introducing the
classical Poisson $s\ell (2)$ algebra as:
$$\eqalign{ s\ell (2) = \bigl\{ J_3 , J_+ , J_- \bigr\} \quad ; \quad &
\bigl\{ J_3 , J_{\pm} \bigr\} = \pm J_{\pm}\cr
& \bigl\{ J_+ , J_- \bigr\} = -2J_3 }\eqno\eq$$
we now prove the

\noindent\underbar{{\bf Proposition 4.30}}:  {\it The classical Poisson--
bracket algebra corresponding to (4.23) is isomorphic to the
subalgebra of} $U \bigl( s\ell (2)\bigr) = {\cal G}$:
$${\cal G} \equiv \bigl\{ J_3^a \, J_+^{\epsilon + b/2} \, J_-^{b/2}
\,;  a\in
N; \, (b, \epsilon ) \in Q\bigr\}\eqno\eq$$

\par\underbar{{\bf Proof}}.  From the definition
in (4.30), we compute directly the Poisson
bracket of two generators of ${\cal G}$ as:
$$\eqalign{ \bigl\{ J_3^{a_1} \, J_+^{\epsilon_1 + b_{1/2}} J_-
^{b_{1/2}} \, , &  J_3^{a_2} \, J_+^{\epsilon_2 + b_{2/2}} \, J_-
^{b_{2/2}}\bigr\}\cr
& = \bigl( a_1 (\epsilon_2 + {b_2\over 2} ) - a_1 \, {b_2\over 2}
\bigr) \, J_3^{a_1 + a_2 - 1} \, J_+^{\epsilon_{1+1} + {b_1+ b_2\over
2}} J_- ^{{b_1 + b_2\over 2}} - (1 \leftrightarrow 2)\cr
& + 2 (\epsilon_1 + {b_1\over 2} ) ({b_2\over 2} ) \, J_3^{a_1 + a_2 +
1} \, J_+ ^{\epsilon_1 \epsilon_2 + {b_1 + b_2 - 2\over 2}} \, J_-
^{{b_1 + b_2\over 2}} - (1\leftrightarrow 2)\cr
& = (a_1 \epsilon_2 - a_2 \epsilon_1 ) J_3^{a_1 + a_2 -1} \, J_+
^{\epsilon_1 + \epsilon_2 -  {b_1 + b_2\over 2}} \, J_- ^{{b_1 +
b_2\over 2}}\cr
& + (\epsilon_1 b_2 - \epsilon_2 b_1 ) \, J_3^{a_1 + a_2 + 1} \, J_+
^{\epsilon_1 + \epsilon_2 + {b_1 + b_2 - 2\over 2}} \, J_- ^{{b_1 +
b_2 - 2\over 2}}\quad  {\bf ::} }$$
The occurence of $S\ell (2)$ is not surprising in a problem concerning a
potential $v(x) = \mu x^2 + g/x^2$, which has an associated $S\ell (2) $
symmetry [DFF].  The energy of an eigen--operator is understandably
obtained as the difference between the exponents of $J^+$ and $J^-$
while the spin is identified as $a+ \epsilon + b \geq 0$.  One
notices that the set of exponents allowed for $J^+$ and $J^-$ is
asymmetrical since $b/2 \in {1\over 2} N$ and $\epsilon + b/2 =
{3b\over 2} , {3b\over 2} - 2\cdots - {b\over 2}$.  The redundancy of
indices is associated here to the existence of the Casimir operator
$J_3^2 - J^+ J^-$ which all but reduces (4.30) to the 2-index symmetric
algebra.

{\bf 5. \underbar{Limits of the Algebra}}

It is finally interesting to study, in the light of the results in Chapter 2,
the two monomial limits of the potential $v(x) = \mu x^2 + g/x^2$.

a)  \underbar{$\mu\rightarrow 0 $}  \quad The potential becomes a
``non--integrable" monomial according to Theorem 2.4.  We recall that
the energy eigenvalues are here normalized by a factor $2\sqrt{\mu},$
hence they become $0$ and the theorem is valid (it only
mentions non--vanishing eigenvalues).

b)  \underbar{$g\rightarrow 0$}  \quad The potential becomes $\mu
x^2$.  In this case, it is clear from (4.10) that $a$ and $b$ are
totally redundant variables and the only meaningful quantity is $(a+b)$.
Accordingly the algebra (4.23) reduces to a 2-index algebra:
$$\bigl[ B^{(b_1)} (\epsilon_1 ) , B^{(b_2)} (\epsilon_2 )\bigr] =
(b_2 \, \epsilon_1 - b_1 \, \epsilon_2 ) B^{(b_1 +b_2 - 1)}
(\epsilon_1 + \epsilon_2 )\eqno\eq$$
Reinterpreting the indices $b_i$ and the energies $\epsilon_i$ leads
to understanding this algebra as the integer-spin subalgebra of the
full eigen--operator algebra described in (2.15).  Due to the fact that
the form (4.10) automatically leads to even powers of $x$ for $a,b$
integers, one cannot obtain the half-integer spin subset.  In order to
get it, one should in particular allow half-integer indices for the
hierarchy Hamiltonians $h_{\pm}^{(n)}$.  Although formally correct, this
approach however induces infinite series of powers of $\alpha$ when $g\not=
0$, and thus goes beyond our restricted definition of eigen--operators
inside the $w_{\infty}$-algebra, and beyond the span of this present
study.
\endpage

\noindent {\bf NOTE ADDED:}

We have noticed that from Propositions 4.24 and 4.30, we can in fact
identify the algebra (4.25):

\noindent
\underbar{{\bf Proposition 4.32:}} \quad {\it The algebra (4.25) is
the classical limit of the extended} $W_\infty (c)$ {\it algebras
(for} $c=4\mu g$ {\it )}.
$$\eqno\eq$$

\noindent
These extended algebras were constructed in [PRS2] precisely as
sub--algebras of the enveloping algebra of $sl(2)$ quotiented by
the ideal generated by $J_3^2- :J_+ J_- : =c$.
Their classical limit is identified as a sub-algebra of the symplectic
algebra on an $sl(2)$ -- coadjoint orbit defined by the quadratic
equation $z^2-xy=4\mu g$.

This orbit is either a cone ($g=0$, corresponding to a pure $w_\infty$
algebra as we have seen) or a 2--sheet hyperboloid ($\mu g <0$) or a
1--sheet hyperboloid ($\mu g > 0$). It may therefore be that the most
relevant underlying fundamental algebra for string theory be not the
particular cone--symplectomorphism algebra $w_\infty$, but the more
general quadric--symplectomorphism algebras $w_\infty (c)$.

\vskip 0.5 truein

\noindent{\bf\underbar{References}}

\item{[Ar]}  V. I. Arnol'd, ``Mathematical Methods of Classical
Mechanics", edited by Springer Verlag, New York (1978).

\item{[AJ1]}  J. Avan, A. Jevicki, ``Classical Integrability and
Higher Symmetries of Collective String Field Theory", {\it Phys.
Lett.} {\bf B 266}, 35 (1991).

\item{[AJ2]}  J. Avan, A. Jevicki, ``Quantum Integrability and Exact
Eigenstates of the Collective String Field Theory", {\it Phys. Lett.}
{\bf B 272}, 17 (1990).

\item{[AJ3]}  J. Avan, A. Jevicki, ``String Field Actions from
$W_{\infty}$", {\it Modern Phys. Lett.} {\bf A7}, 357 (1992).

\item{[Ba1]}  I. Bakas, ``The Large-N Limit of Extended Conformal
Symmetries", {\it Phys. Lett.} {\bf B228}, 57 (1989).

\item{[Ba2]}  I. Bakas, ``The Structure of the $W^{\infty}$-Algebra",
{\it Comm. Mat. Phys.} {\bf 134}, 487 (1989).

\item{[BIPZ]}  E. Br\'ezin, Cl. Itzykson, G. Parisi, J. B. Zuber,
``Planar Diagrams", {\it Comm. Mat. Phys.} {\bf 39}, 35 (1978).

\item{[BK]}  E. Br\'ezin, V. A. Kazakov, ``Exactly Solvable Field
Theories of Closed Strings", {\it Phys. Lett.}  {\bf B236}, 144
(1990).

\item{[Ca]}  F. Calogero, ``Solution of the One--Dimensional $n$-Body
Problem with Quadratic and/or Inversely Quadratic Pair Potentials",
{\it Journ. Mat. Phys.} {\bf 12}, 419 (1971).

\item{[Co]}  S. Coleman, ``1/N", SLAC-Publ. 2484 (1979),
{\it Erice Subnuclear},  {\bf 11} (1979).

\item{[DDGM]}  S. R. Das, A. Dhar, G. Mandal, S. Wadia, ``Gauge Theory
Formulation of the c=1 Matrix Model", Preprint ETH 91-30 (Zurich)
(1991).

\item{[DJe]}  S. R. Das, A. Jevicki, ``String Field Theory and
Physical Interpretation of d=1 Strings", {\it Mod. Phys. Lett.} {\bf A
5}, 1639 (1990).

\item{[DFF]} V. de Alfaro, S. Fubini, G.Furlan, ``Conformal Invariance
in Quantum Mechanics", {\it Nuov. Cim.} {\bf 34}, 569 (1976).

\item{[DJR1]}  K. Demeterfi, A. Jevicki, J. Rodrigues, ``Perturbative
Results of Collective String Field Theory", {\it Modern Phys. Lett.}
{\bf A6}, 3199 (1991).

\item{[DJR2]}  K. Demeterfi, A. Jevicki, J. Rodrigues, "Scattering
Amplitudes and Loop Corrections in String Field Theory", {\it Nucl
Phys.} {\bf B362}, 173 and {\bf 365}, 499 (1991).

\item{[DK]}  P. DiFrancesco, D. Kutasov, ``Correlation Functions
in 2--d String Theory", {\it Phys. Lett. } {\bf B261}, 385 (1991).

\item{[DS]}  M. Douglas, S. Shenker, ``Strings in Less Than One
Dimension", {\it Nucl. Phys.}  {\bf B335}, 635 (1990).

\item{[Dr]}  V. G. Drinfel'd, ``Quantum Groups", Proceeding I.C. M.
Berkeley, {\bf 1}, 798 (1986).

\item{[F]}  L. D. Faddeev, ``Integrable Models in 1+1-Dimensional
Quantum Field Theory", in ``Recent Developments in Field Theory and
Statistical Mechanics", edited by J. B. Zuber and R. Stora, {\it
Elsevier} (1984).

\item{[GK]}  D. Gross, I. Klebanov, ``Fermionic String Field Theory
of c=1 2d Quantum Gravity", {\it Nucl Phys.}
{\bf B352}, 671 (1991).

\item{[GKN]}  D. Gross, I. Klebanov, M. J. Newman, ``The 2-Point
Correlation Function of the 1-d Matrix Model", {\it Nucl. Phys.} {\bf
B350}, 621 (1991).

\item{[GM]}  D. Gross, A. A. Migdal, ``A Non--Perturbative Treatment
of 2-d Quantum Gravity", {\it Nucl. Phys.}  {\bf B340},
333 (1990).

\item{[GMi]} D. Gross, N. Miljkovic, ``A Non-Perturbative Solution of d=1
String Theory", {\it Phys. Lett}  {\bf B238}, 217 (1990).

\item{[Je]}  A. Jevicki, ``Non-Perturbative Collective Field Theory",
Brown HET-807 (1991), to appear in Nucl. Phys. B.

\item{[JS]}  A. Jevicki, B. Sakita, ``The Quantum Collective Field
Method and its Application to the Planar Limit", {\it Nucl. Phys.}
{\bf B165}, 511 (1980).

\item{[Ji]}  M. Jimbo, ``A $q$-Difference Analogue of $U(G)$ and the
Yang-Baxter Equation", {\it Lett. Math. Phys.}  {\bf 11}, 247 (1986).

\item{[JM]}  M. Jimbo, T. Miwa, ``Solitons and Infinite--Dimensional
Lie algebras", {\it Publ. RIMS Kyoto Univ.}  {\bf 19}, 943 (1983).

\item{[KP0]}  I Klebanov, A. M. Polyakov, ``Interaction of Discrete
States in 2d String Theory", {\it Mod. Phys. Lett.}  {\bf A6}, 3273
(1991).

\item{[Li]}  J.  Liouville,  ``Note sur l'int\'egration des \'equations
diff\'erentielles de la dynamique", {\it Journ. de Math}. (Liouville)
{\bf 20}, 137 (1855).

\item{[MPZ]}  D. Minic, J. Polchinski, Z. Zhang, ``Translation-
invariant Backgrounds in 1+1-Dimensional String Theory",  Preprint
UTTG-16-9 (Texas) (1991).

\item{[Mo]}  G. Moore, ``Double Scaled Field Theory at c=1", Preprint
RU-91-12 (Rutgers) (February 1991).

\item{[MS]}  G. Moore, N. Seiberg, ``From Loops to Fields in Quantum
Gravity", Preprint RU-91-29 (Rutgers) (1991).

\item{[No]}  M. Nomura, ``A Soluble Non--linear Bose Field as a
Dynamical Manifestation of Symmetric Group Characters and Young
Diagrams", {\it Phys. Lett.} {\bf B117}, 289 (1986).

\item{[OP]}  M. A. Olshanelsky, A. M. Perelomov, ``Classical
Integrable Finite--Dimensional Systems Related to Lie Algebras", {\it
Phys. Rep.}  {\bf 71}, 313 (1981).

\item{[Pe]}  A. M. Perelomov, ``Exact Results for One-Dimensional
Many-Particle Systems", {\it Sov. Journ. Part. Nucl.}  {\bf 10}, 336
(1979).

\item{[P]}  J. Polchinski, ``Classical Limit of 1+1 Dimensional String
Theory", {\it Nucl. Phys.} {\bf B362}, 25 (1991).

\item{[Po]}  A. Polyakov, ``Singular States in 2d Quantum Gravity",
Preprint PUPT (Princeton) 1289 (1991);
lectures given at Jerusalem Winter School,
1991.

\item{[PRS]}  C. Pope, L.Romans, X. Shen, ``The complete structure of
$W_{\infty}$", {\it Phys. Lett.}  {\bf B236}, 173 (1990).

\item{[PRS2]} C. Pope, X. Shen, L. Romans, ``$W_\infty$ and the
Racah-Wigner Algebra", {\it Nucl. Phys.} {\bf B339}, 191 (1990).

\item{[SFT]}  E. K. Sklyanin, L. D. Faddeev, L. A. Takhtadzyan, ``The
Quantum Method of the Inverse Problem", {\it Theor. Mat. Phys.}  {\bf
40}, 688 (1980).

\item{[Wi]}  E. Witten, ``Ground Ring in 2d String Theory", Preprint IASSNS-
HEP-91-51 (Princeton I.A.S.) (1991); E. Witten, B. Zwiebach,
``Algebraic Structures and Differential Geometry in 2-d String
Theory", Preprint IASSNS-HEP-92-4 (1992).

\end